# DuctApe: a suite for the analysis and correlation of genomic and OmniLog[TM] Phenotype Microarray data

Marco Galardini[1*], Alessio Mengoni[1], Emanuele G. Biondi[2], Roberto Semeraro[1], Alessandro Florio[3], Marco Bazzicalupo[1], Anna Benedetti[3], Stefano Mocali[4]

[1]Department of Biology, University of Florence, Florence, Italy
[2]Interdisciplinary Research Institute USR3078, CNRS-Université Lille Nord de France, Villeneuve d'Ascq, France
[3]Consiglio per la Ricerca e la sperimentazione in Agricoltura, Centro di Ricerca per lo studio delle Relazioni tra Pianta e Suolo (CRA-RPS), Rome, Italy
[4]Consiglio per la Ricerca e la sperimentazione in Agricoltura, Centro di Ricerca per l'Agrobiologia e la Pedologia (CRA-ABP), Florence, Italy

## Abstract
Addressing the functionality of genomes is one of the most important and challenging tasks of today's biology. In particular the ability to link genotypes to corresponding phenotypes is of interest in the reconstruction and biotechnological manipulation of metabolic pathways. Over the last years, the OmniLog™ Phenotype Microarray (PM) technology has been used to address many specific issues related to the metabolic functionality of microorganisms. However, computational tools that could directly link PM data with the gene(s) of interest followed by the extraction of information on gene-phenotype correlation are still missing.
Here we present DuctApe, a suite that allows the analysis of both genomic sequences and PM data, to find metabolic differences among PM experiments and to correlate them with KEGG pathways and gene presence/absence patterns. As example, an application of the program to four bacterial datasets is presented.
The source code and tutorials are available at http://combogenomics.github.io/DuctApe/.

# 1. Introduction

The correlation of complete genotypes (genomes) to complete phenotypes (phenomes) is one of the most challenging tasks of today's biology. Several successful studies have showed that this correlation can be exploited either by experimental or computational approaches [1-3], with important consequences for theoretical and applicative biology. The fundamental starting point of these studies was the definition of a comprehensive picture of both the genome and the phenome of an organism.

The determination of a genomic sequence and its annotation has become a routine task, thanks to the recent advances in both sequencing and annotation technologies [4]. Nevertheless the challenge remains the prediction of the metabolic functions expressed by the genome. In fact, a large fraction of the genes from genomic sequencing efforts has no ascribed function, and even genes with ascribed functions are primarily based on DNA sequence homology, with little or no direct experimental data [5]. Several methods have been proposed and are currently used to construct comprehensive cellular models, such as those applied in the KEGG, MetaCyc and SEED databases [6-8]; all these methods can be integrated with experimental data (i.e. transcriptomics, proteomics, metabolomics) to refine and expand such cellular models [9].

Even though such techniques are of great help, the final expression of the genomic information is represented by the phenotype. In order to assess a rapid functional and phenotypic profiling under different metabolic conditions, a high-throughput approach has been developed by BIOLOG Inc. (Hayward, CA): the Phenotype Microarray (PM) technology. The PM technology, based on the OmniLog™ platform, is a system designed for metabolic and antibiotic resistance assays of both bacterial and eukaryotic cells in 96-well PM plates, comprising more than 2000 temperature-controlled conditions. In each well the metabolism of the cell can be monitored using respiration as a reporter system. More specifically, if the cell has an active metabolism, there would be a flow of electrons to NADH, which determines the reduction of a tetrazolium dye and the consequent production of a purple colour [10]. An automatic camera then records the colour optical density every 15 minutes, providing data over the course of several days: these data are directly stored in a computer and can be analyzed by means of the OmniLog™ PM software [11]. Several other software have been developed to enhance the tools supplied with the OmniLog™ instrument, such as the PhD database [12], RetroSpect™[13] and PheMaDB [14]. Such applications are mainly customized databases which have been designed for the management and the storage of phenotypic data obtained from the OmniLog™ PM software. This last software displays the PM measurements as 96-wells plate layout and provides the kinetic parameter values from each curve, but without considering the entire curve shapes and kinetics. In fact, as recently highlighted by Vaas and co-workers (2012), the PM respiration kinetics contains additional valuable biological information, which needs to be better exploited. In their work the authors proposed a software solution (in the form of the R package '*opm*') for exploiting multiple respiration curves from PM data and providing also a detailed statistical estimation of the results [15, 16]. Although the visualization of the PM raw data was significantly improved as compared to the native OmniLog™ PM software, no method to link the results to putative metabolic pathways was proposed. Even though several works have been published to assess the link between genomic and PM data [17-20] or attempted to improve genome annotation [21], a completely automated approach is still missing.

Here we propose an easy-to-use software suite called "DuctApe", which analyses and links together both the genomic and the phenomic data and suggests genetic explanations of metabolic phenotypes. The suite uses the KEGG database as a source of information for metabolic pathways, divided in reactions (that are the proteins in each genome) and compounds (many of which are directly mapped to PM plates); DuctApe acts by mapping the genome content and the compounds into the same metabolic maps. It also provides various network statistics to help predict which parts of the metabolic network may be more related to the utilization of a specific compound. The suite flexibility allows the analysis of several kinds of experimental setups: i) a single strain experiment, ii) mutational experiments with one reference strain and one or more mutants, and iii) pangenomic experiments, having more organisms simultaneously. The various components and algorithms of the DuctApe suite are presented here, together with an application to four datasets: i) four strains of the symbiotic nitrogen-fixing model bacterium *Sinorhizobium meliloti*, whose phenotype has been determined by using the entire bacterial PM plates set and compared to their genomes [22]; ii) four *Acinetobacter* strains for which both the complete bacterial PM plates set and the genomes were available [19]; iii) a complete bacterial PM plates set on *Zymomonas mobilis* strain ZM4 [23], and iv) the genomic sequences of 15 *Escherichia coli* strains.

## 2. Materials and methods

The tool has been developed as a "suite", which is constituted by three individual modules: 1) dgenome: for genomic data analysis; 2) dphenome: for PM data analysis; 3) dape: for genomics and phenomics combined analysis; the third module is also used to setup the experiment and handle the KEGG data. Each module acts on a project file (more specifically a SQLite database) that is portable and which contains all the data needed to perform the analysis.

### 2.1 dgenome

The dgenome module is used to handle the genomic data, and to return the metabolism reconstruction according to the KEGG database. In this reconstruction each protein that has a KEGG annotation is mapped to its KEGG reaction ID and the corresponding pathway. The user can directly provide the list of KEGG identifiers (i.e. the KEGG orthology IDs given by the KAAS automatic annotation server [24]) or a local KEGG database to which the proteins of each organism will be queried through a Blast-BBH, using the Blast+ software [25]. Once that a list of KEGG identifiers is available, the metabolic network is reconstructed using the KEGG REST API (REpresentational State Transfer Application Programming Interface); more specifically, each KEGG orthology ID is used to retrieve the corresponding reaction IDs, the RPAIR IDs [26], and the relative pathway IDs in which each fetched reaction exercises its function. The '*main*' RPAIR IDs (Reactant PAIRs) are used to map the reactions and compounds to the metabolic maps.

If the genomic data from more than one organism is used (pangenomic experiment), the information on the conserved (core), accessory and unique metabolisms can be extracted through the detection of all the orthologs (the so-called pangenome); the user can import the orthologs classification as provided by an external algorithm, like InParanoid [27] or use the internal Blast-BBH approach [28]. The provided Blast-BBH algorithm is implemented in a way that allows the efficient use of computers with many CPUs, in order to reduce the computation time. The protein sequences are handled by the BioPython package [29].

The orthologs defined by the dgenome module can also be used to improve the annotation of genes to the KEGG database, given the notion that orthologous genes should share the same function. Proteins with no KEGG annotation which are orthologs to annotated proteins inherit the same KEGG annotation; the program flags the proteins as 'automatically annotated by DuctApe'. This approach can be then used to improve, at least partially, the automatic KAAS annotation, increasing the number of genes mapped to the KEGG pathways.

Once each protein has been mapped to the KEGG pathways, the module provides as output a series of statistics about the number of proteins mapped to KEGG for each genome, either as the number proteins with a KEGG orthology ID, reactions and the number of distinct reaction IDs (representing the distinct metabolic abilities of each organism). Since more than one protein in a genome can be mapped to the same reaction ID (having therefore the same function), the number of distinct reaction IDs is always lower than the number of mapped reactions. If more than one organism is present in the project file (either as mutants or pangenome), the number of reaction IDs exclusive to each organism is provided; in case of a pangenome project, the number of exclusive reaction IDs is provided for each pangenome fraction (core, dispensable, accessory and unique) (Figure 1). The number of exclusive reaction IDs represent the unique metabolic features of one organism and may provide a first indication on the genetic basis of the phenotypic variability exploited by the dphenome module.

### 2.2 dphenome

This dphenome module allows the user to provide phenomic data in the form of PM output files, which are then analysed and mapped to the metabolic pathways by means of the KEGG compounds database.

In order to facilitate the analysis and interpretation of the PM experiment, here we propose the use of the "ActiVity index" (AV) as a single and concise parameter to rank and compare each respiration curve; using a single parameter provides both qualitative (presence/absence) and quantitative information about the ability to effectively metabolize a compound or to growth in a specific culture condition. The AV parameter is obtained through the application of a k-means clustering algorithm on five normalized curve parameters, specifically the length of the lag phase, the slope of the curve, the average height of the curve, the maximum cell respiration and the area under the curve. This clusterization ensures that curves with similar shapes are grouped together; the AV value is obtained by ranking the clusters by their average area under the curve. Therefore an AV value equal to zero indicates a curve with no metabolic activity, while higher AV values will be assigned to curves with increasing levels of metabolic activity (Figure 2).

The user can use as input PM data as csv (comma-separated values) files provided by the OmniLog™ system, or as YAML/JSON files, as provided by the *opm* package [16] or other DuctApe projects. One single experiment or one or more replica can be provided in the same file. By default the module supports the twenty bacterial PM plates (PM1-20), but any other kind of 96-wells plate can be imported by the user, thus extending the module scope. The raw data is then processed by the application of the following steps:

1. Parsing
2. Control signal subtraction (optional)
3. Signal refinement
4. Parameters extraction
5. AV calculation
6. Replica management
7. Growth curves plots

The input data is parsed by the module and stored inside the project file (1); the user can then decide if the control signal has to be subtracted on the plates that have one or more control wells, or if a "blank" plate (with no inoculants) should be used for this purpose. This step can be avoided, as some concerns regarding the control signal subtraction have been recently issued [15](2). The PM curve parameters are extracted through the fit on one of three sigmoid functions (Logistic, Gompertz, Richards), reformulated to facilitate the parameters extraction [30]. In order to avoid incorrect fittings or even failures due to spurious peaks in some curves (see for example the curves in Figure 3), the growth curve signal is smoothed through the application of a Blackman window [31], as implemented in the NumPy package [32](3). The actual fit is performed using the SciPy package (through the curve_fit function) [33]. If none of the three functions can be fitted to the PM curve, up to two more rounds of signal smoothing are applied; if the curve fit cannot be applied even after three rounds of smoothing, the plateau, slope and lag parameters are set to zero (4). The user can also decide to import the curve parameters from other software, such as *opm*, providing the PM data in YAML/JSON format.

The AV has a value between 0 and a user defined value k (k > 0), which indicates those wells that exhibit low or none metabolic activity (0) and those with higher metabolic activity (k) inside that particular experiment. The AV parameter is calculated through a k-means clusterization (with k clusters) on five growth curve parameters (max, area, average height, lag time, slope). The MeanShift clustering algorithm is also applied on the PM data, with the addition of a PM plate in which each well has a value of zero for each parameter, in order to verify that there is a real clusterization of the growth curves in distinct clusters, as this algorithm has no fixed amount of clusters. If only one cluster is returned by the MeanShift algorithm the AV of all wells is set to zero. The clusterization is performed using the Scikit-learn package [34](5).

Since in many PM experiments one or more replica of each assay are used, there may be also the need to highlight those growth curves whose shape is significantly different from the average: the dphenome module can search for those replica that are distant from the average AV more than an user-defined AV delta and flag them as 'purged', meaning that they are not considered in the other analysis. The program has five replica purging policies, to help the user decide which PM curves have to be maintained. The user can decide to remove a specific replica ('replica'), to keep the curves with higher or lower AV value ('keep-max' and 'keep-min'), or to keep just the replica with highest or lowest AV value ('keep-max-one' and 'keep-min-one') (6).

The PM data can also be represented as curves plots, created by the Matplotlib package [35]. These curve plots, compared to those generated by the machine vendor software, allow the comparison of more than two strains at the same time, with plate-wise plots, single wells plots and plate-wise AV heatmaps (Figure 4). Additionally it is also possible to display the entire metabolic profile of one or more strains in a ring view, called the 'Activity ring'. Such graphical representation can report either the total AV values (Figure 5a) or the differences in the AV value between any selected strain versus the others by the application of the 'diff mode' (Figure 5b). The differences are represented by colour intensities (blue and orange stripes indicate higher and lower values than the selected reference strain, respectively). These visualizations allow an easier detection of the metabolic categories which are more variable among the tested strains (7).

For those PM compounds that are mapped to the KEGG database, the module uses the KEGG REST API to map them to their corresponding pathway; the ID of those reactions and RPAIRs that use each compound as substrate are retrieved as well, in order to facilitate the combined metabolic analysis.

The module outputs a series of tables with the AV value and the curve parameters for each curve, as well as a series of summary plots; the PM data is also exported in YAML/JSON format to be further analyzed by other PM analysis software, such as *opm* [16].

### 2.3 dape

The "dape" - contraction of DuctApe - module is used to combine the data gathered by the other two modules and provide insights into the genetic determinants of the observed phenotypic variability, using the KEGG database as the metabolic information source.

Relevant information about all the KEGG pathways (reactions, compounds, RPAIRs) are retrieved through the KEGG REST API; each provided genome is then mapped to the pathways through the reactions fetched by the dgenome module, while the information regarding the PM compounds (including AV values) are retrieved from the dphenome module.

The dape module uses these two sources of information, according to the experiment type (single, mutants, pangenome), to rank each pathway/PM compound, thus facilitating the recognition of those pathways putatively related to phenotypic features. For 'single organism' experiments the pathways are ranked by the number of mapped reactions, while the PM compounds are ranked by their average AV. For 'mutants' experiments the pathways are ranked by the number of mapped mutated reactions, while the PM compounds are ranked using the difference in the AV value with respect to the wild-type. In case of a 'pangenome' experiment, the pathways and PM compounds are ranked by their genetic and phenotypic variability, respectively. The genetic variability ($\Gamma$) of each pathway (p) is defined as follows:

$$\Gamma_p = \frac{\delta_p}{\tau_p}$$

where $\delta_p$ indicates the number of variable reactions over the number of distinct reactions mapped in pathway p ($\tau_p$); a "variable" reaction is defined as a reaction differentially present in each strain. A value of one for this parameter indicates those pathways whose gene content is completely variable between the strains in the pangenome, while a value of zero indicates those pathways whose mapped reactions are conserved in all the provided genomes.

The phenotypic variability ($Y$) for each PM compound (c) is defined as follows:

$$Y_c = \frac{2}{n(n-1)} \sum_{\substack{k,j=1\ldots n \\ k<j}} |AV_j - AV_k|$$

where *n* is the number of strains in the pangenome, and $AV_j$ and $AV_k$ are the average AV found for the compound c in organism j and k, respectively. Compounds that show no variability at the metabolic levels, as measured by the PM assay, have phenotypic variability equal to zero. Conversely a higher value for this parameter indicates those compounds whose metabolic effects are significantly variable between the tested strains. Since many compounds can be mapped to one or more pathways, those pathways with both high genetic variability and containing compounds that induce a high phenotypic variability represent hotspots for the metabolic network variability.

The reactions and compounds content of each pathway of the metabolic network can be analyzed by looking at the interactive metabolic maps generated by the dape module, in which specific colour codes are used to highlight reactions presence/absence and the AV as measured by the PM experiment (Figure 6). For instance, in the case of a 'pangenome' experiment, the reactions will be coloured with blue if present in all strains (conserved), and orange if present only in a subset of strains

(variable). Using these maps it is therefore possible to identify the single reactions (and relative genes) that may be directly related to a specific PM compound metabolism.

Moreover, thanks to the NetworkX [36] module, several statistics can be computed for the whole reconstructed metabolic network and for each pathway, like the number of mapped reactions, the number and length of connected components (that are in fact distinct and independent metabolic modules) and the average AV for each pathway. Each pathway is also converted in a graph, in which each node represents a compound and each edge a reaction; the dape module saves such graph reconstructions in gml (Graph Modelling Language) format for further analysis on graphs visualization software, like Gephi [37].

## 3. Results

The functionalities and performances of the three DuctApe modules were tested using four case-studies: i) four strains belonging to the *Sinorhizobium meliloti* species (genomes and PM data) [17, 22], ii) four strains belonging to the *Acinetobacter* genus (genomes and PM data) [19], iii) *Zymomonas mobilis* strain ZM4 (genome and PM data) [23] and iv) 15 strains belonging to the *Escherichia coli* species (genomes) (Supplementary material S1). The following sections describe the application of each DuctApe module on the above-mentioned datasets. More specifically, the dgenome Blast-BBH algorithm has been tested on the three datasets having multiple strains (*S. meliloti*, *Acinetobacter* and *E. coli*); the other DuctApe features have been tested on the three datasets for which both genomic and PM data were available (*S. meliloti*, *Acinetobacter* and *Z. mobilis*).

### 3.1 dgenome

The pangenome of the *S. meliloti*, *Acinetobacter* and *E. coli* datasets has been computed using the dgenome Blast-BBH algorithm (e-value threshold $1e^{-10}$ and BLOSUM80 matrix) and compared with the pangenome computed using OrthoMCL (default parameters and BLOSUM80 matrix) [38] and the combination of InParanoid and MultiParanoid (default parameters and BLOSUM80 matrix) [27, 39] (Supplementary material S1). The three methods estimated a very similar core genome size for the three datasets: about 3000 orthologous groups (OGs) for *E. coli* and *Acinetobacter*, and about 5000 OGs for *S. meliloti* (Figure 7a), thus confirming the accuracy of the BBH algorithm implemented in the dgenome module. A broader difference was observed for the accessory and unique genome compartment (Figure 7b and 5c, respectively), where the Blast-BBH algorithm predicted a higher number of OGs, especially for the *E. coli* dataset. This difference is due to the ability of OrthoMCL and InParanoid to detect paralogs, thus reducing the number of OGs in the dispensable genome. As recently pointed out, Blast-BBH may be unable to correctly detect orthologs in species with many gene duplications: for such cases, the user should then import orthology predictions from algorithms that are able to detect paralogs [40].

The proteins belonging to the *Acinetobacter*, *S. meliloti* and *Z. mobilis* datasets were mapped to KEGG using the KAAS annotation web server [24]. Concerning the *S. meliloti* dataset, a similar number of proteins (ca. 3000) were mapped to KEGG orthologs in each of the four genomes (Figure 1a), with an almost identical number of distinct reactions IDs. When looking at the pangenome level, the highest fraction of proteins mapped to KEGG comes from the core genome (> 50%), with a limited fraction of reactions mapped in the accessory and unique genome (< 25%) (Figure 1b), thus confirming the lower genetic variability found in the metabolic network, as opposed to the overall genetic variability. A similar behaviour has been observed in the *Acinetobacter* dataset, where about half of the proteins of each genome was found to be mapped to KEGG orthologs; as in the *S. meliloti* dataset, the majority of proteins mapped in KEGG belonged to the core genome. The small genome of *Z. mobilis* ZM4 (1736 proteins) was found to have a slightly higher fraction of proteins mapped to KEGG orthology (1063, 61% of the proteins) as compared to the other two datasets (Figure 8).

The pangenomes of the *S. meliloti* and *Acinetobacter* datasets have been used to improve the KEGG annotation of each strain, as described in the "Materials and methods" section. The orthology relationships led to the additional annotation of 279 and 237 proteins in the *S. meliloti* and *Acinetobacter* datasets, respectively, thus demonstrating the ability of the dgenome module to improve the KEGG metabolic reconstruction.

### 3.2 dphenome

In order to evaluate the dphenome module, the entire bacterial PM plates set was used (see Supplementary material S1) for the three datasets for which PM data were available (*S. meliloti*, *Acinetobacter* and *Z. mobilis*). The AV has been calculated using five clusters (k=5) for the *S. meliloti* dataset and six clusters (k=6) for the *Acinetobacter* and *Z. mobilis* datasets. The optimal number of clusters has been chosen through a simple elbow test. For each of the five parameters used in the AV calculation, the sum of squared errors was computed using a number of clusters ranging from 2 to 12; the number of clusters that was responsible for the highest reduction in the error was considered as optimal (Figure 9) [41].

The curve parameters extracted by the dphenome suite on the *S. meliloti* dataset have been also compared with the ones calculated by the *opm* package (which calculates the area under the curve, the lag time, the maximum signal and the slope of the curve), showing a slightly higher variability for the lag and slope parameters between the two software, which is possibly due to the smoothing algorithm applied by dphenome to remove spurious peaks in the raw data. The AV value computed using the *opm* parameters has been compared to the one provided by DuctApe, which also showed a very low variability between the two parameters set, with the exception of plates PM07 and PM08, where DuctApe indicated an higher AV than *opm* (see Figure 10 and Supplementary material S2). The biggest difference in the AV values in these two plates could be explained by stochastic effects rather than to the plates themselves, as there is no relevant difference in comparison with the curves in the other plates. For some curves (in plates PM07 and PM08) we observed a difference in the clusterization between the two programs, possibly due to the overall small differences in the parameters calculation. These small differences could results in a curve being assigned to an adjacent cluster when calculating the AV, therefore having a difference of 1 AV unit. Two examples of this phenomenon are presented in Figure 3.

Comparing the high/low metabolic activity for each PM compound, we have found that in the *S. meliloti* dataset 80.8% of the substrates used were shared among the four *S. meliloti* strains (difference of AV < 2, indicating only a moderate variability in the PM curve shape), whereas the 19.2% of the metabolic repertoire was unique for each of them. The highest number of unique "more active" metabolic features was detected in PM plates inoculated with AK58 and BL225C strain (67 and 61, respectively), suggesting a higher metabolic potential of these strains as compared to the others. In particular, the BL225C strain exhibited "less active" phenotypic traits in 12 conditions only, as shown in Table 1. In contrast, the Rm1021 strain seems to be the less metabolically active strain, as it showed a lower number of "more active" metabolisms (16) and the highest number of "less active" metabolisms (119). When looking at the single PM compound categories (Table 2), trends highlighted in Table 1 were confirmed for each category (with the sole exception of phosphate and sulphur compounds), with strains AK83 and Rm1021 having the lowest proportion of active compounds when compared to the other two strains. In particular, strain AK83 performed poorly on carbon and nitrogen sources, on nitrogen sources from peptides and in the various osmolarity and pH assays, while strain Rm1021 had the lowest activity on nutrient stimulants and was also found to exhibit the lowest resistance to chemical agents. This result was expected, since strain Rm1021 has been cultured in laboratory conditions for many generations. Most of the overall phenotypic variability seems to be concentrated on the carbon sources, with almost 20% of the tested compounds showing a high phenotypic variability (difference of AV >= 2) (Table 3). The same information could be retrieved using the 'Activity ring' visualization (Figure 5), which can also be visualized using the 'diff mode', in order to visually detect any difference in higher/lower metabolic reaction between the reference strain Rm1021 and the other strains (Figure 2b). A significant increase of blue colour intensity appeared in the "Nitrogen (peptides)" and the "Chemicals" categories, confirming the lower metabolic activity of Rm1021 as compared to the other *S. meliloti* strains.

The analysis on the *Acinetobacter* phenome confirmed the earlier published reports [19]. The clinical isolate *A. baumannii* strain 19606 was found to be the most metabolically versatile strain, either by looking at the number of "more active" metabolic features (48 compounds, Table 1) or by the proportion of active compounds for each category (Table 2): strain 19606 showed higher proportion of active compounds for carbon and nitrogen sources, on nitrogen sources from peptides and in the resistance to chemical agents. When compared to *A. calcoaceticus* strain RUH2202, strain 19606 was found to be more active in all PM categories with the sole exception of phosphate and sulphur sources and, on a limited extent, in the presence of nutrient stimulants, thus confirming previous reports [19].

Also the analysis on the *Z. mobilis* ZM4 phenome confirmed earlier published reports [23]. As reported in Table 2, *Z. mobilis* was found to have a dramatically reduced ability to actively metabolize carbon sources (only 3.6%), when compared to the more versatile *S. meliloti* and *Acinetobacter* strains. All the compounds indicated by Bochner et al. as positive growth conditions or chemical agents that could inhibit strain ZM4 growth were confirmed using the dphenome module. In addition, other compounds that favoured or inhibited an active metabolism of strain ZM4 were exclusively highlighted by dphenome; those compounds showed a similar curve shape as the compounds highlighted by Bochner and collaborators (see Figure 11 for an example) [23], thus suggesting that the automatic approach followed by dphenome is able to retrieve most of the relevant metabolic features. The same analysis has been conducted using the AV calculated from the parameters computed by the opm package (Supplementary material S8); opm correctly highlighted most of the growth conditions indicated by Bochner et al. (269 over 313), while 44 growth conditions were not retrieved.

### 3.3 dape

The dape module was used to highlight those metabolic pathways with both high genetic ($\Gamma_p$) and phenotypic variability ($Y_c$) and to inspect those KEGG pathways to highlight the single reactions responsible for the metabolic variability. Concerning the *S. meliloti* dataset, the KEGG pathway that describes the nitrogen metabolism (map00910) was found to have the highest genetic variability (0.36) and a number of PM compounds with the highest phenotypic variability (i.e. 2.5 for L-Asparagine and L-Glutamine as nitrogen sources); considered the ecological importance of the nitrogen metabolism for the nitrogen-fixer *S. meliloti*, this metabolic map has been therefore analyzed in detail (Figure 6a). Even though most of the reactions are catalyzed by proteins encoded by genes which were not retrieved from the genomic data (white boxes), some reactions appeared to be controlled by genes of the core genome (blue). More importantly, three orange reactions indicated the presence of the following functional genes located on the dispensable genome: nitrite reductase (reaction 1.7.2.1), nitrous-oxide reductase (reaction 1.7.2.4), and nitric oxide reductase (reaction 1.7.2.5). This indicates that such functions are not shared between all strains: in fact those genes encoding for these enzymes were not detected in AK58 and AK83 strains. Similarly, most of the metabolic substrates involved in the nitrogen cycle are not included into the PM microplates set (white circles) but several substrates could be used (or not) by the four strains (coloured circles). For instance, all the samples showed active metabolism for some specific substrate (i.e. Ammonia, L-Glutamate, etc.) as well as no active metabolism for some other substrate (i.e. Hydroxylamine, Formamide, etc.) which are indicated with a grey circle on the map. The red border of circles related to some substrates (i.e. Nitrite, Nitrate, etc.) indicated that the four strains differently used those specific substrates and that the AV values differed for more than 2 AV units. The simultaneous visualization of both genomic and phenomic information allowed to link the presence of nitrite reductase (NO-forming reaction 1.7.2.1), nitrous-oxide reductase (reaction 1.7.2.4) and nitric oxide reductase (reaction 1.7.2.5) with the inability to actively metabolize nitrite as a nitrogen source in Rm1021 and BL225C strains, at least under these experimental conditions. It is known in fact that Rm1021 strain has a functional denitrification pathway [42] as well as other *S. meliloti* strains [43], while in strain AK58, which is completely lacking this pathway (reaction 1.7.2.1, reaction 1.7.2.4 and reaction 1.7.2.5), the metabolism is clearly directed toward the production of ammonia. Therefore in strains Rm1021 and BL225C the denitrification pathway allows the use of nitrite as input for the irreversible production of gaseous nitrogen, possibly reducing the nitrogen pool available for the rest of the metabolism.

The dape analysis on the *Acinetobacter* dataset confirmed previous published reports. The compound showing the highest phenotypic variability was found to be D-Glucarate as a carbon source (3.33) and it was found in the ascorbate and aldarate metabolism (map00053). This pathway was found to have one of the highest values of genetic variability (0.67). The reactions involved in the metabolism of D-Glucarate into 2-Oxoglucarate were found to be missing in strain RUH2202 and RUH2624, which also showed no active metabolism on this compound as a carbon source (Figure 6b). Specifically the missing genes were a glucarate dehydratase (reaction 4.2.1.40) and a deoxyketoglucarate dehydratase (reaction 4.2.1.41). The presence of D-glucarate in the human body and the ability of the pathogen strain 19606 to actively metabolize this compound as an input for the cytrate cycle confirms earlier reports [44-46], thus demonstrating the ability of the dape module to automatically highlight relevant metabolic hotspots.

The application of dape to the *Z. mobilis* dataset correctly highlighted the genes responsible for the metabolism of D-Fructose as a carbon source. This compound is one of the few carbon sources metabolized by this strain and it is found inside the fructose and mannose pathway (map00051). Two genes responsible for the Fructose phosphorylation were predicted, specifically a hexokinase (reaction 2.7.1.1) and a fructokinase (reaction 2.7.1.4) (Figure 6c). Previous reports confirm that these genes are involved in the metabolism of D-Fructose in *Zymomonas mobilis* [47], again confirming the predicting value of the dape module.

These examples show how the dape module can automatically highlight the metabolic variability hotspots, thus facilitating the analysis on both genomic and phenomic experiments.

## 4. Discussion

We propose DuctApe as a software solution for the visualization of PM data, the exploration of both the genomic information and the phenotypic expression of microbial cells, providing also comprehensive insights into the cellular metabolism.

The dgenome module implements functionalities that were already available as separate tools in other software, such as KAAS [24] for the metabolic reconstruction and InParanoid [27] for the orthology algorithm; the strong innovation introduced by this module is the ability to easily obtain and organize data, as well as making them available to the dape module for the final metabolic network reconstruction. The pangenome construction algorithm is designed to be parallelizable, allowing a faster analysis in systems with many CPUs available. The pangenome reconstruction is also used to improve the KEGG metabolism reconstruction through the annotation transfer inside orthologous groups.

The dphenome module represents an important novelty of the DuctApe suite. This tool provides an automated classification and identification of the PM curves and applies the AV as a unique comprehensive kinetic parameter for the computational high-throughput processing of the raw data. The AV value here proposed is a single and reliable parameter that incorporates the information deriving from the curve shapes, going far beyond the mere presence/absence paradigm or the partial evaluation of the metabolic reactions based on the parametric analysis tool of the native OmniLog$^{TM}$ PM software. In fact, such parametric analysis method is based only on few data points of the curve shapes, thus consistently reducing the biological information content of each experiment, as recently reported [15]. The AV value is based on the combination of parameters that are calculated by the measurement of all data points of the curve. In fact, for a comprehensive comparison of the PM curves several parameters have to be considered to derive a meaningful biological estimation. Although a raw combination of few parameters of a curve into a single one is expected to introduce biases, we took into account that the combination of 5 parameters could reduce the load of the "correlated" parameters (i.e. area and average height) and at the same time emphasize the biological information. Therefore, the AV can be considered as a useful and reliable measure able to automatically calculate the metabolic activity of a large number of samples, replacing the application of more sophisticated multivariate data analysis. The use of a single value with a definite range to discriminate the metabolic activity inside the experiment allows an easier analysis and an easier embedding of the phenotypic data inside the metabolic maps. Quick and informative visualization of the AV values of the selected strains can be easily achieved using the 'ring' visualization provided in Figure 5 while, at the same time, retaining the possibility of getting deeper insights into more detailed metabolic categories and substrates. This kind of graphical representation is particularly suitable when facing huge and complicated datasets such as the ones obtained from the entire PM plate set on two or more bacterial strains.

The dape module represents the most important innovation of the software suite and, to the best of our knowledge, it represents the first tool that allows the user to link and visually analyse both genomic and phenomic high-throughput datasets. Following a study in which genomes belonging to *Bacillus cereus* were linked to PM data by comparing their carbohydrate metabolic pathways [48], DuctApe now allows the user to automatically perform this task for all the metabolic pathways present in the KEGG database, together with the analysis on the whole metabolic network reconstruction though the calculation of metabolic networks statistics.

The idea of using the KEGG database and the development of such tool came from the native OmniLog™ PM software which enables the user to identify a single substrate of any PM microplate linking the displayed picture to the KEGG database. However, the dape module improves such link, allowing the users to simultaneously assess any metabolic feature of all the substrates of the entire PM plate set and, moreover, to integrate this information with the genomic data. This kind of data processing will most probably enhance data modelling of genome-wide metabolic pathway and gene annotation.

Although each module of the DuctApe suite can be also used independently from each other, their integration and the use of the dape module allows the user to get into a first reliable biological evaluation of the results. Further work is still necessary to optimize the integration of genomic and phenomic data; in fact, as DuctApe is based on the KEGG platform, the number of the possible genomic explanations of a specific phenotypic feature is strongly limited, either by the KEGG database itself (i.e. the available microbial metabolic pathways are basically obtained by few model-strains, such as *E. coli*), and by the number of substrates which are accessible on PM plates. Furthermore it is well known that the phenotypic diversity of a cell is also affected by mechanisms as, for example, regulatory and signal transduction systems and membrane transporters, whose functionality could not be directly detected by PM system and KEGG-based genome annotation. Future implementation including also transcriptome, proteome and metabolome data could then be implemented to allow more detailed analyses on genome/phenotype relationships.

## Availability

Source code and a detailed online manual is available in the project website (http://combogenomics.github.io/DuctApe/); all the data discussed here can be reproduced using the data provided in the Supplementary material S1, or by downloading the ductape_data git repository (https://github.com/combogenomics/ductape_data/releases/paper). The parameters calculation comparison between DuctApe and the *opm* package can be replicated using the data provided in the Supplementary material S2, or by downloading the ductape_vs_opm git repository (https://github.com/combogenomics/ductape_vs_opm/releases/paper).

## Aknowledgements

We are grateful to Dr. Bochner and Dr. Ziman for having provided the *Z. mobilis* dataset, with Prof. Peleg and Dr. Dijkshoorn for having provided the *Acinetobacter* dataset and to Emanuele Bosi for the parapy script included in the ductape_data repository. We are also grateful to three anonymous reviewers for having greatly improved the manuscript with their insightful comments.

*Funding*: This work was supported by the Italian Ministry of Research (PRIN 2008 research grant contract No.TCKNJL, "Il pangenoma di *Sinorhizobium meliloti*: L'uso della genomica per il miglioramento agronomico dell'erba medica") and by the project MARKERINBIO (DM 8348, 5/27/2010).The funders had no role in study design, data collection and analysis, decision to publish, or preparation of the manuscript.

## Supplementary material

**Supplementary Material S1.** ductape_data repository. Input genomes and PM data are provided, along with a runner script to reproduce the data shown in the manuscript.

.**Supplementary Material S2.** ductape_vs_opm repository. Input PM data and runner scripts are provided, in order to reproduce the comparison with *opm* discussed in the manuscript.

**Table 1.** Unique metabolic functions*

|  | *S. meliloti* | | | | *Acinetobacter* | | | |
|---|---|---|---|---|---|---|---|---|
|  | AK58 | AK83 | BL225C | Rm1021 | 19606 | RUH2202 | RUH2624 | SH024 |
| More active | 67 | 11 | 61 | 16 | 48 | 1 | 17 | 20 |
| Less active | 60 | 33 | 12 | 119 | 5 | 85 | 1 | 3 |
| Total | 127 | 44 | 73 | 135 | 53 | 86 | 18 | 23 |
| % | 6.1 | 2.3 | 3.8 | 7.0 | 2.8 | 4.5 | 0.9 | 1.2 |

* The number of wells (over 1920 total) for which a single strain shows an exclusive phenotype with respect to the other three.

**Table 2.** Proportion of compounds of each PM category whose AV is equal or over 3 in each analyzed strain.

| Category | *S. meliloti* | | | | *Acinetobacter* | | | | *Z. mobilis* |
|---|---|---|---|---|---|---|---|---|---|
| | AK58 | AK83 | BL225C | Rm1021 | 19606 | RUH2202 | RUH2624 | SH024 | ZM4 |
| Carbon | 32.3 | 29.7 | 38.5 | 34.9 | 30.2 | 13 | 24 | 24.5 | 3.6 |
| Nitrogen | 59.4 | 21.9 | 41.7 | 38.5 | 38.5 | 18.8 | 30.2 | 26 | 18.8 |
| Phosphate and sulphur | 51 | 75 | 77.1 | 55.2 | 0 | 4.2 | 31.3 | 1 | 35.4 |
| Nutrient stimulation | 6.3 | 7.3 | 5.2 | 1 | 0 | 1 | 0 | 4.2 | 1 |
| Nitrogen peptides | 68.8 | 29.9 | 61.8 | 49.3 | 61.1 | 9 | 40.3 | 20.8 | 36.5 |
| Osmolytes and pH | 29.7 | 16.7 | 22.4 | 20.8 | 27.1 | 7.3 | 30.7 | 25 | 49.5 |
| Chemicals | 51.5 | 49.2 | 56.1 | 44.2 | 62.9 | 29.3 | 61.7 | 60.7 | 75.1 |

**Table 3.** Proportion of compounds of each PM category for which the average difference between the strains is equal or over 2 AV (main diffs).

| Category | S. meliloti | Acinetobacter |
|---|---|---|
| Carbon | 18.8 | 4,1 |
| Nitrogen | 5.2 | 3,1 |
| Phosphate and sulfur | 8.3 | 3,1 |
| Nutrient stimulation | 1.0 | 2,1 |
| Nitrogen peptides | 5.2 | 8 |
| Osmolytes and pH | 2.6 | 1,6 |
| Chemicals | 4.4 | 1,3 |

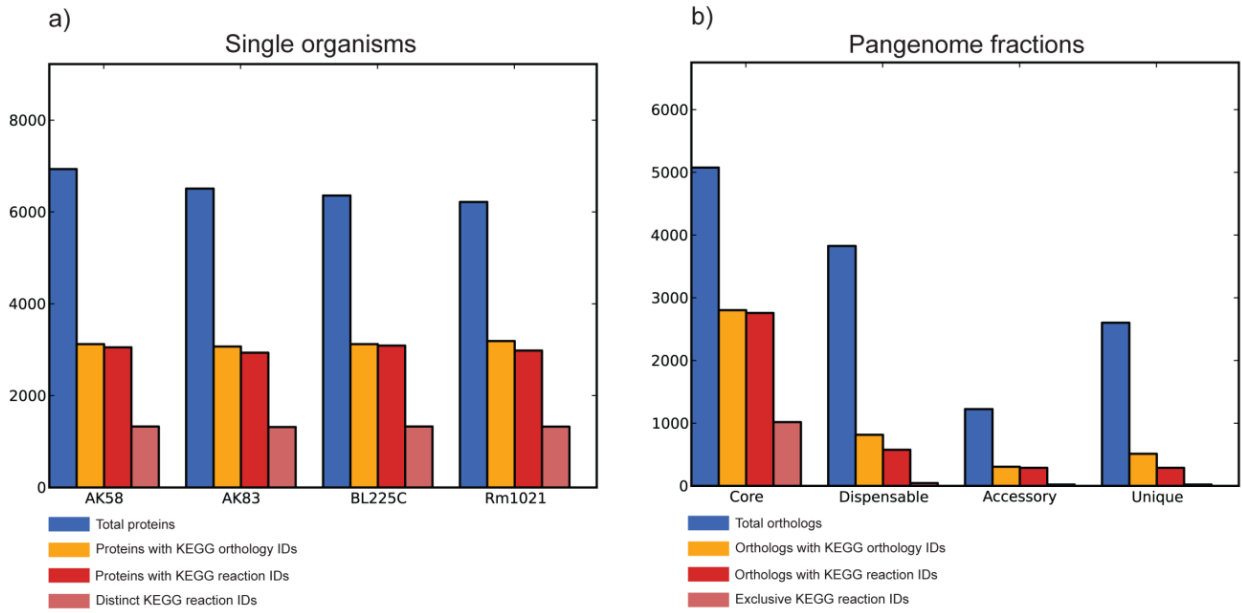

**Figure 1.** dgenome plots on the *S. meliloti* dataset. a) Proteome sizes of the four *S. meliloti* strains and the fractions of proteins mapped to KEGG, KEGG reactions and the number of distinct reaction IDs. b) Pangenome partitions sizes in number of orthologous groups and fraction of orthologs mapped to KEGG, KEGG reactions and the number of exclusive reaction IDs.

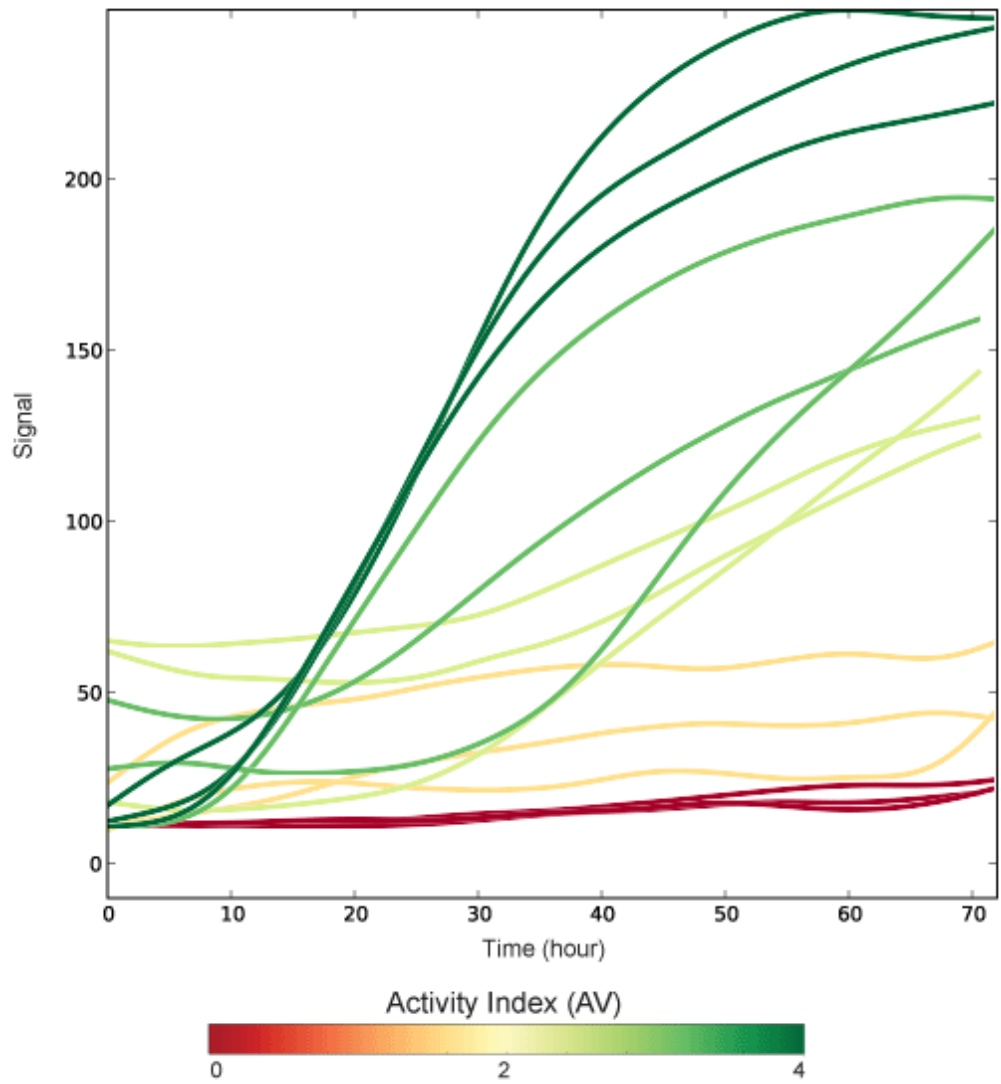

**Figure 2.** Activity index example curves, from the *S. meliloti* dataset. Three random PM curves for each AV cluster are shown.

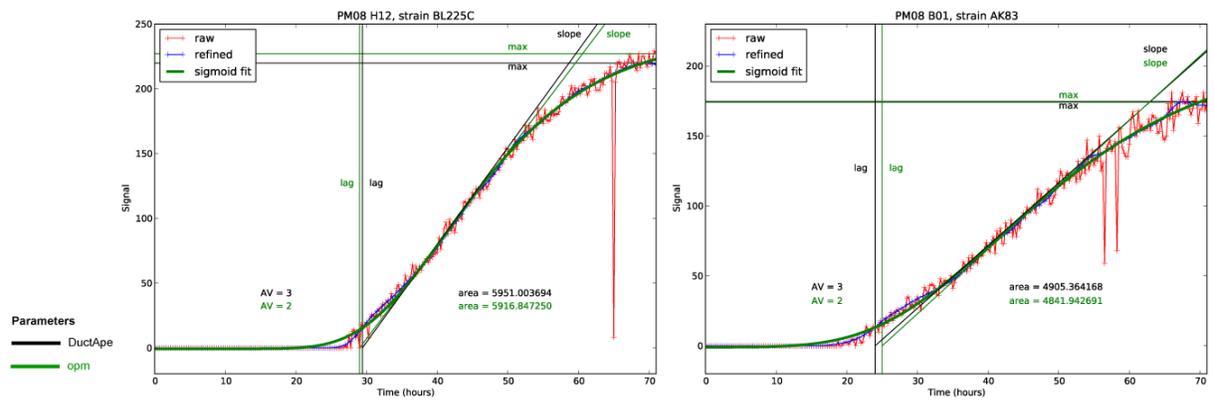

**Figure 3.** Example application of the smoothing algorithm and sigmoid curve fitting on raw PM data. The parameters computed by dphenome and *opm* are compared as well.

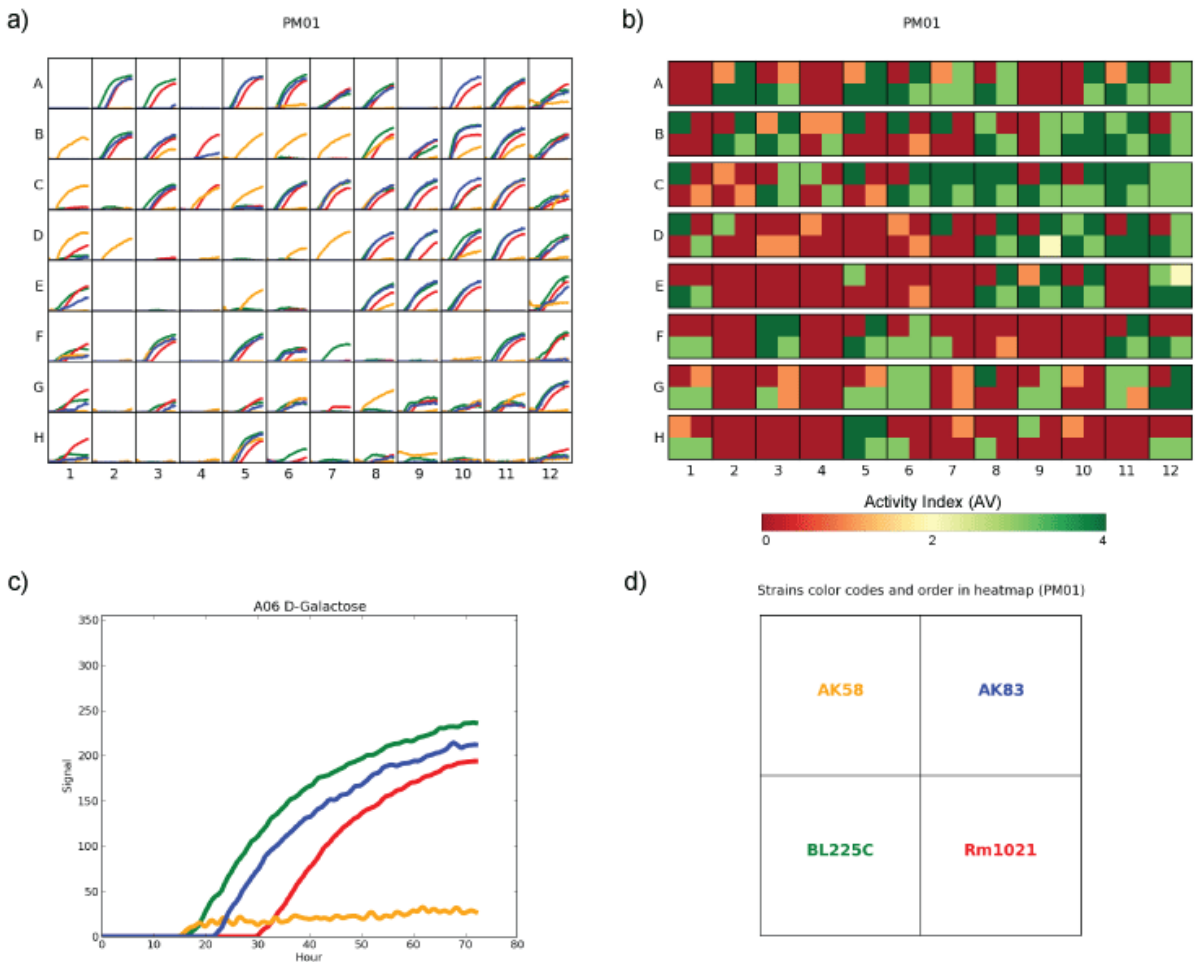

**Figure 4.** dphenome curve plots. a) Overall plate plot; each curve represents PM data from one strain. b) Plate Activity heatmap; for each well, the AV found for each strain is reported. c) Single curve plot; each curve represents PM data from one strain. d) Colour codes for each strain and strains position in the Activity heatmap.

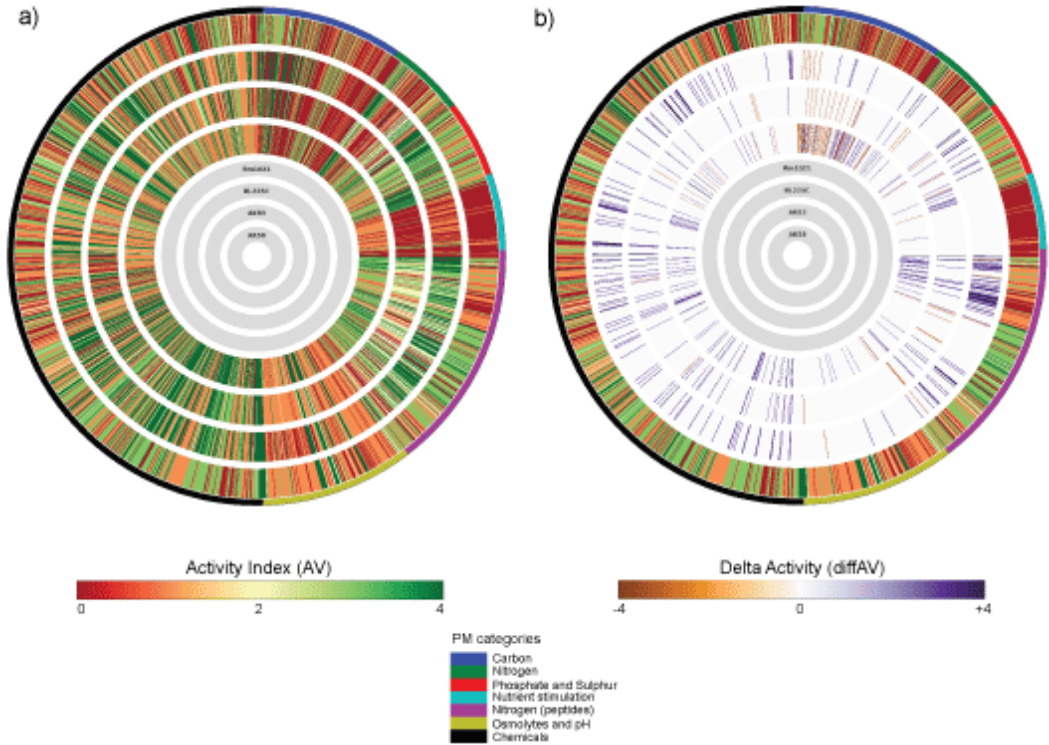

**Figure 5.** Activity rings from the *S. meliloti* PM data; grey inner circles indicate the strains order; external circle indicates the PM categories. a) The Activity Index (AV) calculated for each strain and well is reported as colour stripes going from red (AV=0) to green (AV=4). b) Diff mode: the difference with the AV value of strain Rm1021 is reported when equal or higher of 2 AV, grey otherwise; purple colour indicate an highe activity with respect to Rm1021, orange colour a lower activity.

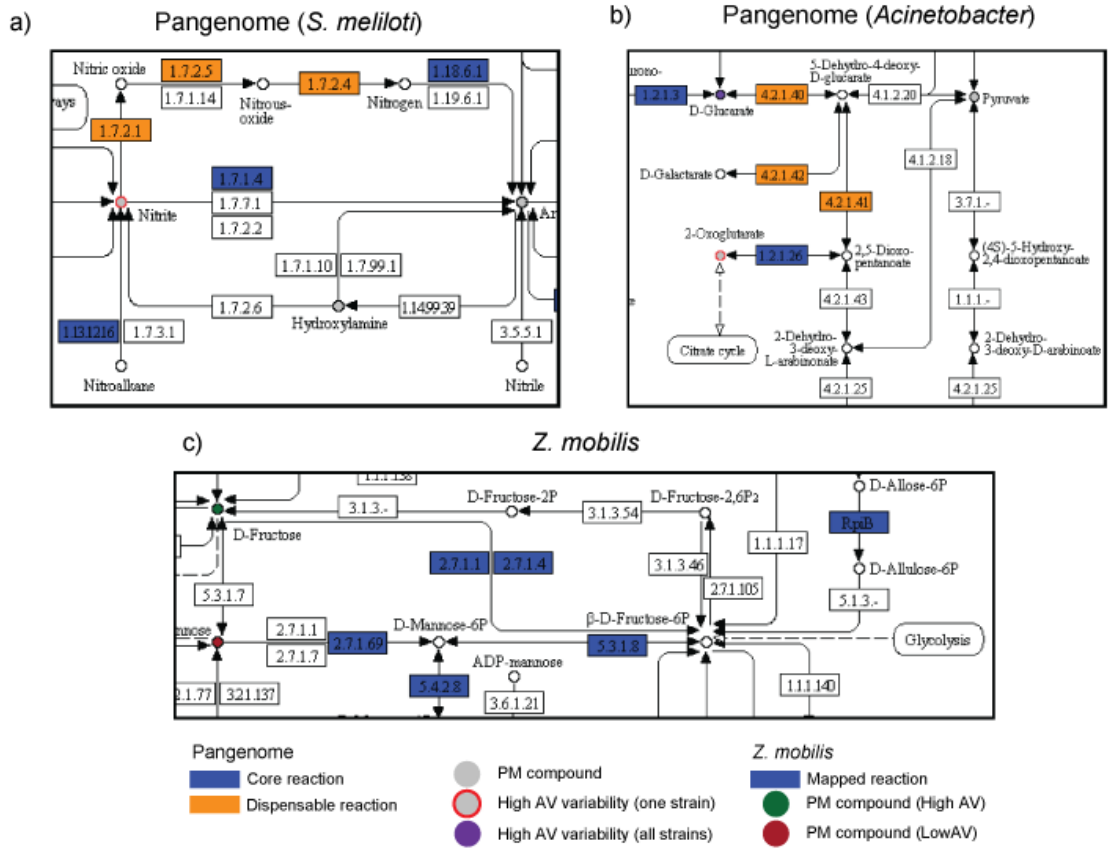

**Figure 6.** Metabolic network analysis using the dape module; boxes represent reactions while circles represent compounds. a) Nitrogen metabolism (map00910) in the *S. meliloti* pangenome: core reactions are coloured in blue, variable reactions are coloured in orange; compounds present in PM plates are filled with grey. Red circles around compounds highlight those compounds for which at least one strain has an AV difference with another strain equal or higher than 2 AV, while compounds coloured in purple indicates a phenotypic variability equal or higher than 2 AV. b) Ascorbate and aldarate metabolism (map00053) in the *Acinetobacter* pangenome; the same colour codes as a) are used. c) Fructose and mannose pathway (map00051) in the *Z. mobilis* genome; reactions mapped in the genome are coloured in blue, while compounds are coloured according to their AV value.

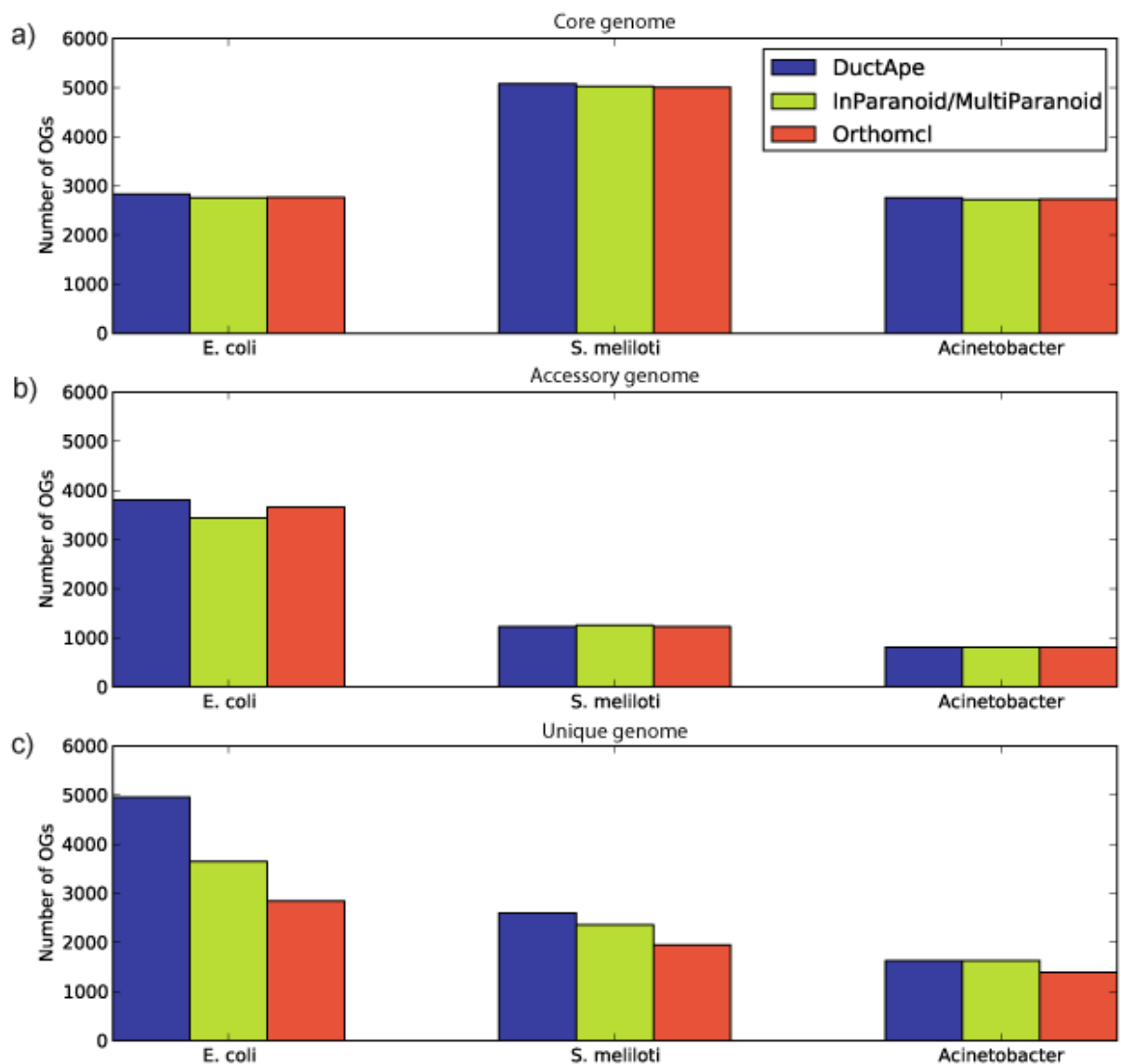

**Figure 7.** Comparison of pangenome prediction performances of DuctApe, OrthoMCL and InParanoid/MultiParanoid in the three pangenomic datasets (*E. coli*, *S. meliloti* and *Acinetobacter*). The number of Orthologous Groups (OGs) belonging to the core (a), accessory (b) and unique (c) genome were compared for the three algorithms.

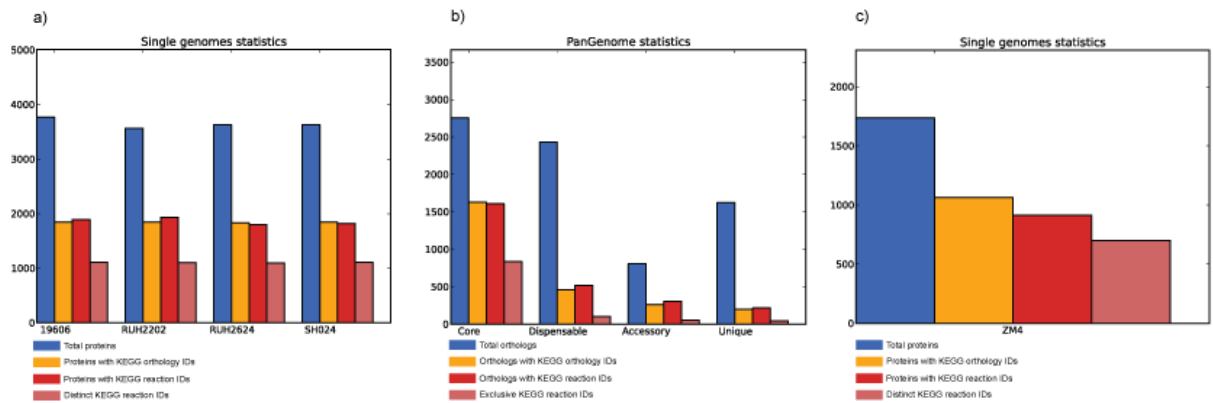

**Figure 8.** dgenome plots of the *Acinetobacter* and *Z. mobilis* datasets. a) Proteome sizes of the four *Acinetobacter* strains and the fractions of proteins mapped to KEGG, KEGG reactions and the number of distinct reaction IDs. b) Pangenome partition sizes in number of orthologous groups and fraction of orthologs mapped to KEGG, KEGG reactions and the number of exclusive reaction IDs. c) Proteome size of the *Zymomonas mobilis* ZM4 genome and the fractions of proteins mapped to KEGG, KEGG reactions and the number of distinct reaction IDs.

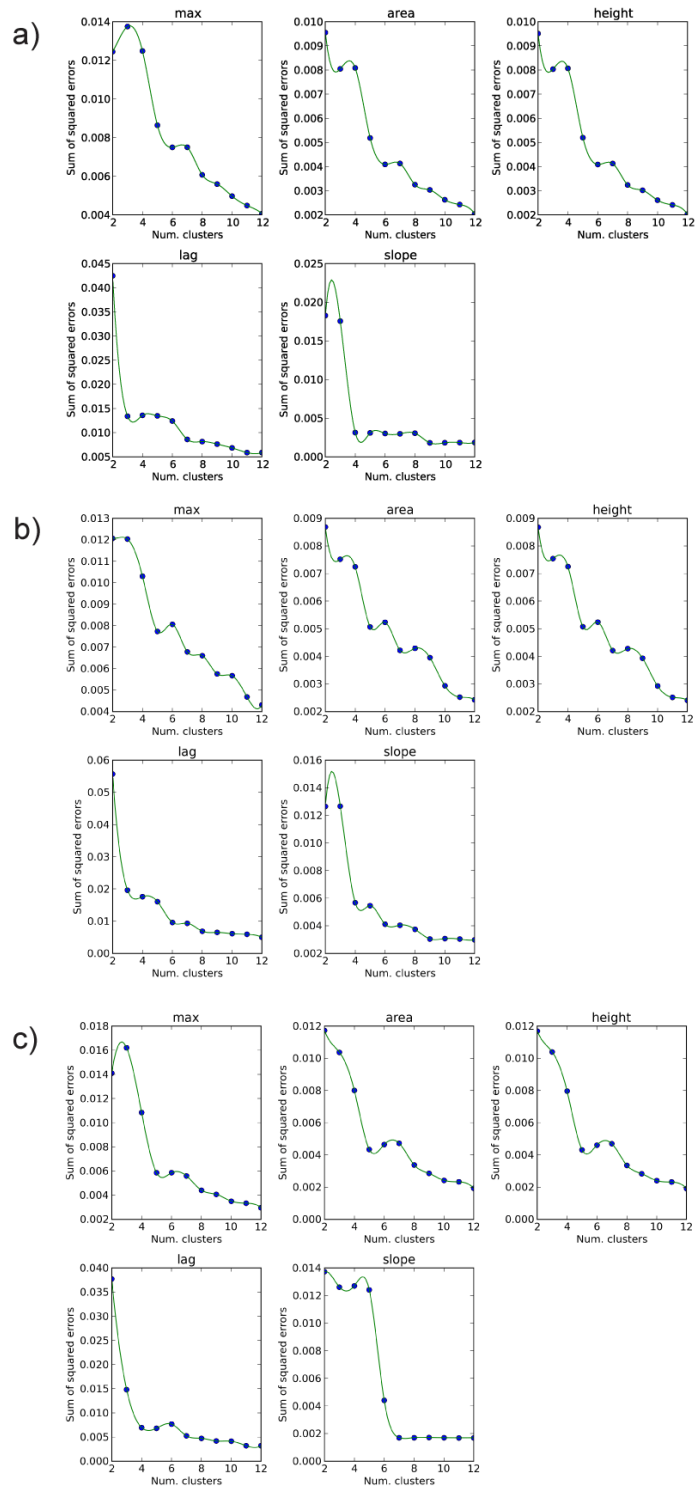

**Figure 9.** dphenome elbow tests. For each parameter, the sum of squared errors is computed for the k-means clusters computed with values of k ranging from 2 to 12. a) Elbow test for the *S. meliloti* dataset. b) Elbow test for the *Acinetobacter* dataset. c) Elbow test for the *Z. mobilis* dataset.

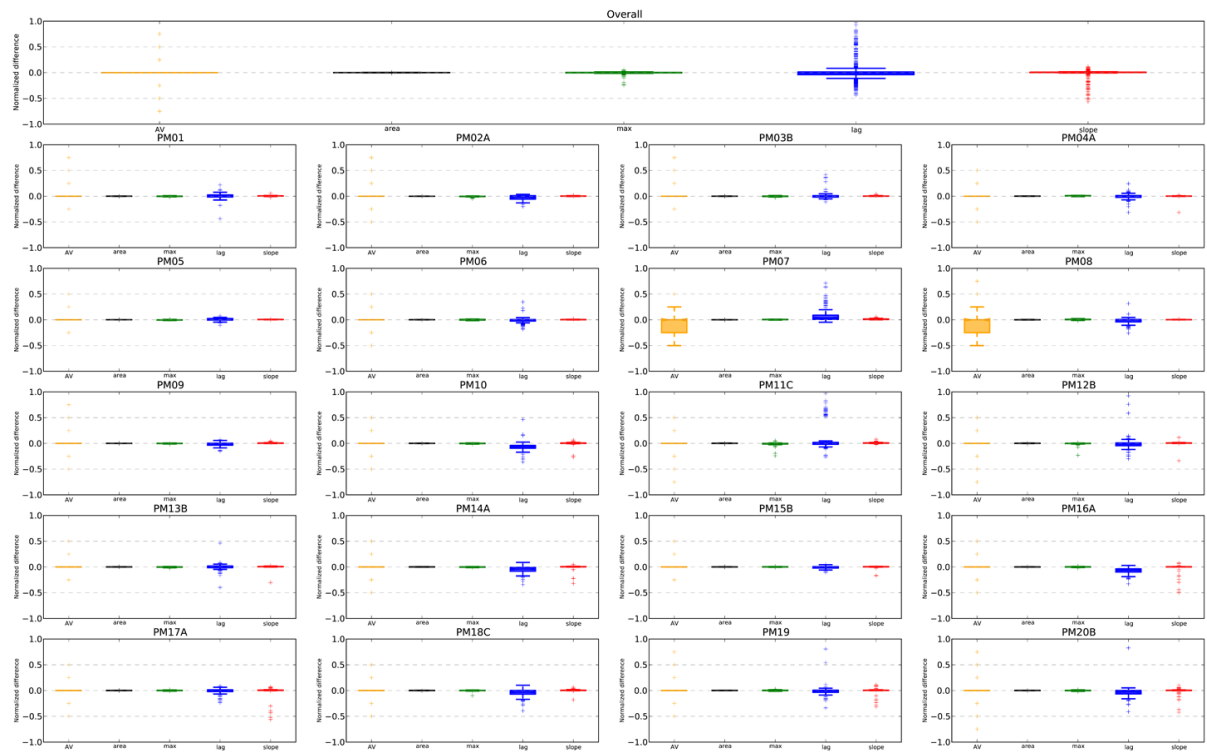

**Figure 10.** Comparison of curve parameters estimation provided by the *opm* package. Overall and per-plate comparisons are provided.

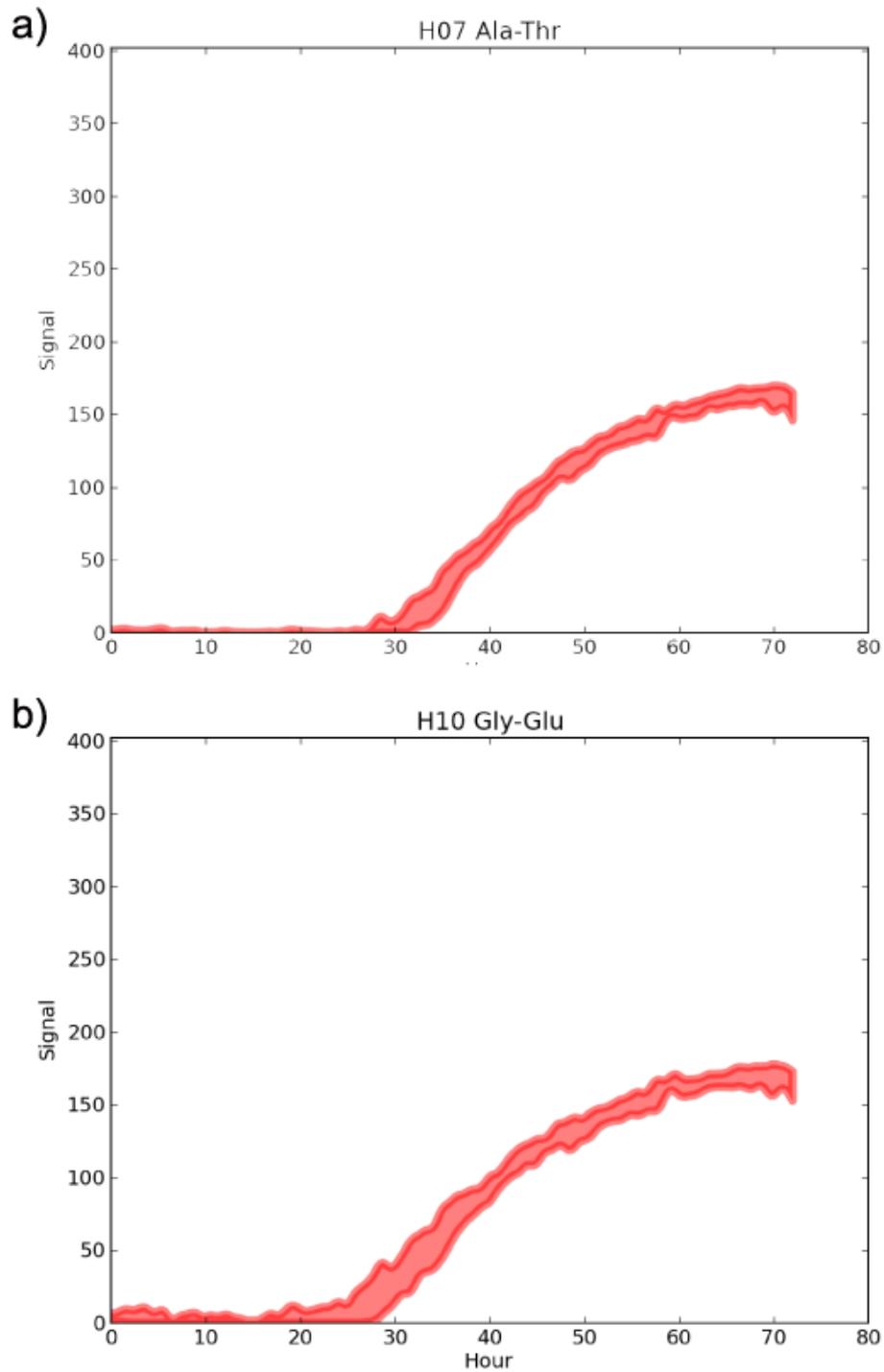

**Figure 11.** PM curves of *Zymomonas mobilis* ZM4 from plate PM03. a) PM curve of a compound highlighted as actively metabolized by dphenome but not in [23]. b) PM curve of a compound highlighted as actively metabolized by both analysis.